\documentclass[12pt]{iopart}

\usepackage{iopams}
\usepackage{graphicx}
\usepackage{amssymb}
\usepackage{ifthen}
\usepackage{amsfonts}

\newcommand{\be}{\begin{equation}}
\newcommand{\ee}{\end{equation}}
\newcommand{\bea}{\begin{eqnarray}}
\newcommand{\eea}{\end{eqnarray}}

\begin{document}

\title{On the convexity of Relativistic Ideal Magnetohydrodynamics}

\author{Jos\'e-Mar\'ia Ib\'a\~nez$^{1,2}$, Isabel Cordero-Carri\'on$^{3,4,5,6}$, 
Miguel-\'Angel Aloy$^1$, Jos\'e-Mar\'ia Mart\'{\i}$^{1,2}$ and Juan-Antonio Miralles$^7$}

\address{$^1$ Departamento de Astronom\'{\i}a y Astrof\'{\i}sica, Universidad de Valencia,
E-46100 Burjassot, Spain}

\address{$^2$ Observatori Astron\`omic, Universidad de Valencia,
  E-46980 Paterna, Spain}

\address{$^3$ Departamento de Matem\'atica Aplicada, Universidad de Valencia,
E-46100 Burjassot, Spain}

\address{$^4$ 
	Laboratoire Univers et Th\'eories (LUTH), 
	Observatoire de Paris/CNRS/Universit\'e Paris Diderot, 
	5 place Jules Janssen, 
	F-92190 Meudon, France
}

\address{$^5$ 
	IFPA,
	D\'ep. AGO, 
	Universit\'e de Li\`ege, B\^at B5, 
	Sart Tilman, B-4000 Li\`ege, Belgium
}

\address{$^6$ 
	Namur Center for Complex Systems (naXys) \& D\'epartement de Math\'ematique,
	Universit\'e de Namur, 
	Rue de Bruxelles 61, 
	B-5000 Namur, Belgium
}
\address{$^7$ Department of Applied Physics, University of Alicante,
Campus de Sant Vicent del Raspeig, E-03080 Alicante, Spain}

\ead{jose.m.ibanez@uv.es, isabel.cordero@uv.es
  miguel.a.aloy@uv.es, jose-maria.marti@uv.es, ja.miralles@ua.es}

\begin{abstract}
We analyze the influence of the magnetic field in the convexity
properties of the relativistic magnetohydrodynamics system of
equations. To this purpose we use the approach of Lax, based on the
analysis of the linearly degenerate/genuinely non-linear nature of the
characteristic fields. Degenerate and non-degenerate states are discussed
separately and the non-relativistic, unmagnetized limits are properly
recovered. The characteristic fields corresponding to the
material and Alfv\'en waves are linearly degenerate and, then, not
affected by the convexity issue. The analysis of the characteristic
fields associated with the magnetosonic waves reveals, however, a
dependence of the convexity condition on the magnetic field. 

The result is expressed in the form of a generalized fundamental
derivative written as the sum of two terms. The first one is the
generalized fundamental derivative in the case of purely
hydrodynamical (relativistic) flow. The second one contains the
effects of the magnetic field. The analysis of this term shows that it
is always positive leading to the remarkable result that the presence
of a magnetic field in the fluid reduces the domain of thermodynamical
states for which the EOS is non-convex. 
\end{abstract}


\submitto{\CQG}

\maketitle

%
\section{Introduction}

 There are many astrophysical scenarios governed by relativistic 
magnetohydrodynamical processes as, e.g., the production of 
relativistic jets emanating from Active Galactic Nuclei, 
the structure and dynamics of pulsar wind nebulae, the mechanisms 
triggering the explosion in core-collapse supernovae, or the 
production of Gamma Ray Bursts. These scenarios are
nowadays the subject of intensive research by means of numerical
simulations thanks to recent advances in numerical relativistic
magnetohydrodynamics (RMHD) that exploit the fact that the RMHD
equations obeying a causal equation of state (EOS) form a hyperbolic
system of conservation laws~\cite{Anile89}.

  Matter at densities higher than nuclear matter density can undergo
first-order phase transitions to various phases of matter, such as
pion condensates~\cite{Kostyuk01}, hyperonic matter ~\cite{Oertel13} 
or deconfined quark matter ~\cite{Gorenstein05,Gorenstein12}. Several
authors~\cite{Harry09,Bombaci12,Peres13} have studied, from different
points of view, the influence that those exotic states of matter at
extreme high densities have on, e.g., the  dynamics of stellar core
collapse supernovae, the evolution of proto-neutron stars, or the
collapse to black hole. 

  The classical Van der Waals (VdW) EOS is a well known example of EOS
displaying a first-order phase transition. Fluids having a
thermodynamics governed by a VdW-like EOS exhibit, outside the region
of the phase transition, non-classical gasdynamic behaviours in a
range of thermodynamic conditions characterized by the negative value
of the so-called fundamental derivative, $\mathcal
G$~\cite{Thompson43, Menikoff89, Guardone10}  
\begin{equation}
{\mathcal G} := - \frac{1}{2} \,V
\,\displaystyle{\frac{\displaystyle{\left.\frac{\partial^2 p}{\partial
          V^2}\right|_s}}{\displaystyle{\left.\frac{\partial
          p}{\partial V}\right|_s}}}
\label{G1}
\end{equation}
\noindent
$p$ being the pressure, $V:= 1/\rho$ the specific volume ($\rho$ is
the rest-mass density) and $s$ the specific entropy. The fundamental
derivative measures the convexity of the isentropes in the $p-V$ plane
and if ${\cal G} > 0$ then the isentropes in the $p-V$ plane are
convex, leading to expansive rarefaction waves (and compressive
shocks) \cite{RZ13}. In a VdW-like EOS, or in general in a non-convex EOS, 
rarefaction waves can change to compressive and shock waves to
expansive depending on the specific thermodynamical state of the
system.  These  non-classical phenomena have been observed experimentally 
and their study is, currently, of interest in many engineering applications 
~\cite{Cinnella07, Cinnella11}.

	Besides this thermodynamical interpretation of convexity, there is
an equivalent definition due to Lax~\cite{Lax57} that connects with
the mathematical properties of the hyperbolic system. According to Lax's approach, 
a hyperbolic system of conservation laws \footnote{The books by LeVeque~\cite{LeVeque92} 
and Toro~\cite{Toro09} are recommendable references for those readers interested 
in the basic theory of hyperbolic systems of conservation laws. The monograph
of \cite{Le02} on finite-volume methods for hyperbolic problems pays special attention
to non-convex flux functions (see their Sects. 13.8.4 -definitions of genuine 
non-linearity and linear degeneracy, and their relationsphip with convexity-, and 16.1
-devoted entirely to the study of scalar conservation laws with non-convex flux 
functions-).} is convex if all its characteristic fields
are either genuinely non-linear or linearly degenerate. A
characteristic field $\lambda$ is said to be genuinely non-linear or
linearly degenerate if, respectively,
\begin{equation}
\label{GNLfieldpm}
{\mathcal P} := \vec{\nabla}_{\bf u} \lambda \cdot {\bf r} \ne 0 ,
\end{equation}
\begin{equation}
\label{GNLfield0}
{\mathcal P} := \vec{\nabla}_{\bf u} \lambda \cdot {\bf r} = 0,
\end{equation}
\noindent
for all ${\bf u}$, where $\vec{\nabla}_{\bf u} \lambda$ is the
gradient of $\lambda({\bf u})$ in the space of conserved variables,
${\bf r}$ is the corresponding eigenvector, and the dot stands for the
inner product in the space of physical states.

  In a non-convex system, non-convexity is associated with those
states ${\bf u}$ for which the factor ${\mathcal P}$ corresponding to a
genuinely non-linear field, Eq.~(\ref{GNLfieldpm}),
 is zero and changes sign in a neighbourhood
of ${\bf u}$.

   A virtue of Lax's approach is that it can be applied to other 
hyperbolic systems in which the convex or non-convex character of 
the dynamics is governed by other ingredients beyond the EOS. 
Among these systems are those of relativistic hydrodynamics (RHD) 
and classical magnetohydrodynamics (MHD). In these two cases, 
the convexity of the system has been characterized with the
sign of a generalized fundamental derivative that includes an extra
term depending of the local speed of sound (in the case of
RHD~\cite{Ibanez13}) and the magnetic field (in the case of
MHD~\cite{Serna14}).

  In this work we use the approach of Lax to characterize, 
from a theoretical point of view, the effects of magnetic fields in 
the convexity properties of the RMHD system of equations as a 
previous step to explore its possible impact in the dynamical 
evolution of different astrophysical scenarios. The result is 
presented in the form of an extended fundamental derivative whose 
sign determines the convex/non-convex character of the RMHD system 
at a given state. Our result recovers the proper non-relativistic 
and unmagnetized limits.

  The paper is organized as follows. In Sect.~2, the equations of RMHD
are introduced as a hyperbolic system of conservation
laws. The transformation between primitive and conserved variables are
explicitly written. In Sect.~3 the characteristic structure of the RMHD
equations is discussed and the analysis  of convexity in
non-degenerate states presented. In Sect.~4 the analysis of convexity
is extended to degenerate states. The non-relativistic, unmagnetized
limits are recovered in Sect.~5. Section~6 includes a short summary
and presents the conclusions. Finally, there is an Appendix that
displays the Jacobian matrices of the RMHD system in quasi-linear
form, necessary for the characteristic analysis of Sect.~3.

\section{The equations of ideal relativistic magnetohydrodynamics}

  Let $J^{\mu}$, $T^{\mu \nu}$ and $^*F^{\mu \nu}$\footnote{Throughout
    this paper, Greek indices will run from 0 to 3, while Roman run
    from 1 to 3, or, respectively, from $t$ to $z$ and from $x$ to
    $z$, in Cartesian coordinates.} be the components of the rest-mass
current density, the energy--momentum tensor and the Maxwell tensor of
an ideal (infinite conductivity) magneto-fluid, respectively  

\begin{equation}
J^\mu = \rho u^\mu
\end{equation}

\begin{equation}
T^{\mu \nu} = \rho h^* u^\mu u^\nu + g^{\mu \nu} p^* - b^\mu b^\nu
\end{equation}

\begin{equation}
^*F^{\mu \nu} = u^\mu b^\nu - u^\nu b^\mu,
\end{equation}

\noindent
where $\rho$ is the proper rest-mass density, $h^* =1 + \epsilon +
p/\rho + b^2/\rho$ is the specific enthalpy including the contribution 
from the magnetic field ($b^2$  stands for $b^\mu b_\mu$), $\epsilon$ 
is the specific internal energy, $p$ is the thermal pressure, $p^* = p + 
b^2/2$ is the total pressure, and  $g^{\mu \nu}$ is the metric of the
space-time where the fluid evolves. Throughout the paper we use units
in which the speed of light is $c=1$ and  the $(4 \pi)^{1/2}$ factor
is absorbed in the definition of the magnetic field. The four-vectors
representing the fluid velocity, $u^\mu$, and the magnetic field
measured in the comoving frame, $b^\mu$, satisfy the conditions
$u^\mu u_\mu = -1$ and $u^\mu b_\mu = 0$.

  The equations of ideal RMHD correspond to the conservation of
rest-mass and energy-momentum, and the Maxwell equations. In a flat
space-time and Cartesian coordinates,  these equations read:

\begin{equation}
\label{cont}
J^\mu_{\,\,\,\,,\mu} = 0
\end{equation}

\begin{equation}
\label{e-mom}
T^{\mu \nu}_{\,\,\,\,\,\,,\mu} = 0
\end{equation}

\begin{equation}
\label{Maxwell}
^*F^{\mu \nu}_{\,\,\,\,\,\,\,\,,\mu} = 0,
\end{equation}

\noindent
where subscript $(\,_{,\mu}\,)$ denotes partial derivative with
respect to the corresponding coordinate, $(t,x,y,z)$, and the standard
Einstein sum convention is assumed.  

  The above system can be written as a system of conservation laws as
follows 

\begin{equation}
\frac{\partial {\bf U}}{\partial t} +
\frac{\partial {\bf F}^{i}}{\partial x^{i}} = 0
\label{e:system}
\end{equation}
\noindent
where ${\bf V} = (\rho, v^i, \epsilon, B^i)^T$ is the set of primitive
variables.  The state vector (the set of conserved variables) ${\bf U}$
and the fluxes, ${\bf F}^i$, are, respectively:

\begin{eqnarray}
{\bf U} & = & \left(\begin{array}{c}
  D    \\
  S^i  \\
  \tau \\
  B^i
\end{array}\right),
\label{state_vector}
\end{eqnarray}

\begin{eqnarray}
{\mathbf F}^i & =& \left(\begin{array}{c}
  D v^i \\
  S^j v^i + p^{*} \delta^{ij} - b^j B^i/W \\
  \tau v^i + p^{*} v^i - b^0 B^i/W \\
  v^i B^k - v^k B^i
\end{array}\right).
\label{flux2}
\end{eqnarray}

  In the preceding equations, $D$, $S^j$ and $\tau$ stand,
respectively, for the rest-mass density, the momentum density of the
magnetized fluid in the $j$-direction, and its total energy density,
all of them measured in the laboratory (i.e., Eulerian) frame:

\begin{equation}
\label{eq:D}
  D = \rho W,
\end{equation}

\begin{equation}
\label{eq:Sj}
  S^i = \rho h^* W^2 v^i - b^0 b^i,
\end{equation}

\begin{equation}
\label{eq:tau}
  \tau = \rho h^* W^2 - p^* - (b^0)^2 - D.
\end{equation}

\noindent
The components of the fluid velocity trivector, $v^i$, as measured in
the laboratory frame, are related with the components of the fluid
four-velocity according to the following expression: $u^\mu = W(1,
v^i)$, where $W$ is the flow Lorentz factor, $W^2=1/(1-v^i v_i)$. 

  The components of the magnetic field four-vector in the comoving frame
and the three vector components $B^i$ measured in the laboratory frame
satisfy the relations: 

\begin{eqnarray}
\label{b0}
  b^0 & = & W v_k B^k, \\
\label{bi}
  b^i & = & \frac{B^i}{W} + b^0 v^i.
\end{eqnarray}

  Finally, the square of the modulus of the magnetic field can be written as

\begin{equation}
  b^2 = \frac{B_k B^k}{W^2} + (v_k B^k)^2.
\label{e:b2}
\end{equation}

  The preceding system must be complemented with the time component of
equation~(\ref{Maxwell}), that becomes the usual divergence
constraint 

\begin{equation}
  \label{eq:divb}
  \frac{\partial B^i}{\partial x^i} = 0.
\end{equation}

  An EOS $p=p(\rho,\varepsilon)$ closes the system. Accordingly, the
(relativistic) sound speed
$a_{s}:=\displaystyle{\sqrt{\left.\frac{\partial p}{\partial
        e}\right|_s}}$, $e$ being the mass-energy density of the fluid
$e=\rho(1+\epsilon)$, satisfies $\displaystyle{h a_{s}^{2} = \chi +
  \frac{p}{\rho^{2}} \, \kappa}$, with $\displaystyle{\chi :=
  \left.\frac{\partial\,p}{\partial\,\rho}\right|_{\varepsilon}}$ and
$\displaystyle{\kappa :=
  \left.\frac{\partial\,p}{\partial\,\varepsilon}\right|_{\rho}}$. 

\section{Characteristic structure of the RMHD equations and analysis
  of convexity in non-degenerate states} \label{s:csrmhde}

  The characteristic information of the system of RMHD
(\ref{e:system}) is contained in the set
of eigenvalues and right eigenvectors
$\{\lambda_{\alpha}, {\bf r }_{\alpha}\}_{\alpha =1}^8$ of $\zeta_k{\bf
{\cal B}}^k$, where ${\bf {\cal B}}^i := \displaystyle{\frac{\partial
    {\bf F}^i}{\partial {\bf U}}}$ are the Jacobian
matrices of the vectors of fluxes along the coordinate directions, and
$\zeta_i$ is an arbitrary unitary 3-vector.

  Since the dependence on ${\bf U}$ of the fluxes ${\bf F}^i$ is
implicit, it is useful to write the Jacobian matrices ${\bf {\cal B}}^i$
in terms of
matrices involving only explicit derivatives with
respect to the primitive variables, ${\bf V}$. If we define ${\bf {\cal
A}}^0 :=
\displaystyle{\frac{\partial {\bf U}}{\partial {\bf V}}}$, and
${\bf {\cal A}}^i :=\displaystyle{ \frac{\partial {\bf
      F}^i}{\partial {\bf V}}}$, then we have that ${\bf {\cal B}}^i =
{\bf {\cal A}}^i ({\bf {\cal A}}^0)^{-1}$. Now, the sets of eigenvalues
and right eigenvectors of the system in conservation form,
$\{\lambda_{\alpha}, {\bf r }_{\alpha}\}_{\alpha =1}^8$, and of
the system in quasi-linear form, $\{\lambda^*_{\alpha},
{\bf r }^*_{\alpha}\}_{\alpha =1}^8$, satisfying $(\zeta_k{\bf {\cal
    A}}^k - \lambda^*_{\alpha} {\bf {\cal A}}^0) {\bf r}^*_{\alpha} =
0$, are related according to $\{\lambda_{\alpha}, {\bf r
}_{\alpha}\}_{\alpha =1}^8 = \{\lambda^*_{\alpha}, {\bf {\cal A}}^0
{\bf r}^*_{\alpha}\}$. Matrices ${\bf {\cal A}}^0$ and $\zeta_k {\bf
  {\cal A}}^k $ are displayed in the Appendix.

  Once the eigenvalues and eigenvectors are known, we can analyze the
convexity of the system studying the expression ${\mathcal P}_{\alpha} =
\vec{\nabla}_{\bf U} \lambda_{\alpha} \cdot {\bf r}_{\alpha}$ (see
the Introduction).  Finally, we can take advantage of the fact that, since
${\bf {\cal A}}^0$ is non-singular, then ${\mathcal
  P}_{\alpha} \neq 0$ if, and only if, ${\mathcal P}_{\alpha}^* :=
\vec{\nabla}_{\bf V}
\lambda_{\alpha} \cdot {\bf r}^*_{\alpha} \neq 0$, and perform the
analysis of convexity in terms of ${\mathcal P}_{\alpha}^*$.

  The eigenvalues $\lambda_{\alpha}$ are the solutions of the following
polynomial expression for $\lambda$

\bea
\lambda a 
\Big({\cal E} a^2 - \mathcal{B}^2\Big)
\Big((b^2 + \rho h a_s^2) a^2 G - W_s^{-2} \rho h a^4 -
  a_s^2 G \mathcal{B}^2\Big) & = & 0, 
\label{caract}
\eea

\noindent
where ${\cal E} := \rho h + b^2$, $W_s^{-2} := 1 - a_s^2$ and 
quantities $a$, $G$ and $\mathcal{B}$ were defined in
ref.~\cite{Anile89}, $a := \phi_\alpha u^\alpha$, $G := \phi_\alpha
\phi^\alpha$, $\mathcal{B} := \phi_\alpha b^\alpha$, being, in our
case, $\phi_\alpha := (-\lambda, \zeta_i)$ the normal to the wavefront
propagating with speed $\lambda$ in the spatial direction given by the
unit vector $\zeta_i$.

  As it is well known, the system of (R)MHD is not-strictly
hyperbolic ~\cite{Brio88}. This means that in some cases, two or more eigenvalues can
be equal leading to well studied cases of degeneracy (see 
refs.~\cite{Anile89,Anton10}, for the relativistic case). In Type I
degeneracy, the magnetic field is normal to the propagation direction
of the wavefront (i.e., $\zeta_k B^k = 0$). In Type II degeneracy,
$\zeta_k B^k \neq 0$, but the eigenvalues associated with, at least,
one Alfv\'en wave and one magnetosonic wave are degenerate. 
Leaving aside the particular cases associated with both degeneracy types, 
that will be discussed later, the following list compiles the roots of 
the characteristic equation
(\ref{caract}), $\lambda_\alpha$ ($=\lambda^*_\alpha$), the right
eigenvectors, ${\bf r}^*_{\alpha}$ \footnote{The expressions of the
eigenvectors have been obtained after tedious algebraic
manipulations. They can be verified by direct substituting in
the eigenvalue equation, $(\zeta_k{\bf {\cal A}}^k 
- \lambda^*_{\alpha} {\bf {\cal A}}^0) {\bf r}^*_{\alpha} = 0$.}, 
and their corresponding scalar products, ${\mathcal P}_{\alpha}^*$ 
\footnote{For the scalar products ${\cal P}_{a_{\pm}}^*$ and 
${\cal P}_{m_{\pm}}^*$, the partial derivatives of the corresponding 
eigenvalues with respect to the primitive variables, ${\bf V}$, have been 
computed by implicit derivation of the characteristic equations for
$\lambda_{a_{\pm}}$ and $\lambda_{f_{\pm}}$, respectively, i.e.,
$\mathcal{A} = 0$ and $\mathcal{N}_4 = 0$ (see below).}, in the
non-degenerate, general case.

\begin{itemize}

\item[i)] $\lambda = \lambda_{\rm null} := 0$.  In
  this case, ${\mathcal P}_{\rm null}^*$ is trivially zero. This eigenvalue is
  spurious and is associated with the fact that although the RMHD
  system (\ref{e:system}) consists of eight conservation equations,
  only seven components of the fluxes are non-trivial. Due to the
  antisymmetric character of the induction equation, the flux of
  $\zeta_k B^k$ in the $\zeta^k$-direction is identically zero.

\item[ii)] $\lambda = \lambda_0 := \zeta_k v^k$ is the eigenvalue
  associated with the material waves. The
  corresponding eigenvector is ${\bf r}^*_0 = (-\kappa, 0^i, \chi,
  0^i)^T$, where $\kappa$ and $\chi$ are thermodynamical derivatives
  defined at the end of the previous Section, and $0^i = 0$ ($i = 1, 2,
  3$). The scalar product is ${\mathcal P}_{0}^* = 0$ and, consequently,
  the characteristic field defined by $\lambda_0$ is linearly
  degenerate.

\item[iii)] $\lambda = \lambda_{a_{\pm}}$ are the roots of the
  second-order polynomial in $\lambda$, $\mathcal{A}$,

\be
  \mathcal{A} := {\cal E} a^2 - \mathcal{B}^2.
\label{e:alf}
\ee
\indent
They define the Alfv\'en waves. Since $\zeta_k B^k \neq 0$, then
$a\neq 0$ and the corresponding eigenvectors are
\bea
  {\bf r}^*_{a_{\pm}}  & = &  (0, r_2^i, 0, r_4^i)^T,
\eea
where $r_2^i  = a_1 B^i + a_2 v^i + a_3 \zeta^i$, 
$r_4^i = W a^{-1} (r_2^i \zeta_k B^k - B^i \zeta_k r_2^k)$. 
The coefficients $a_p
\,(p=1,2,3)$ are such that $v_k r_2^k = 1$, 
$\zeta_k r_2^k = - W a$,
and $B_k r_2^k = - v_k B^k W^2$. The scalar products are
\be
{\cal P}_{a_{\pm}}^* =\Big(\frac{\partial \lambda_{a_{\pm}}}{\partial v^i}\Big)\, r_2^i
+ \Big(\frac{\partial \lambda_{a_{\pm}}}{\partial B^i}\Big)\, r_4^i
\propto \Big(\zeta_k r_2^k + W\, a\, (v_k r_2^k)\Big)
 = 0,
\ee
in agreement with the linearly degenerate character of the Alfv\'en waves.

\item[iv)] The four eigenvalues $\lambda_{f_{\pm}}$,
  $\lambda_{s_{\pm}}$, are the roots of the fourth-order polynomial in
  $\lambda$, $\mathcal{N}_4$,

\be
  \mathcal{N}_4 := (b^2 + \rho h a_s^2) a^2  G  - W_s^{-2} \rho h  a^4
- a_s^2 G \mathcal{B}^2,
\label{e:pol-lambda4}
\ee
associated with the fast and slow magnetosonic wavespeeds, respectively.
Since $\zeta_k B^k \neq 0$, then $a \neq 0$ and the corresponding
eigenvectors are
\be
{\bf r}^*_{m_\pm} = (r_1, r_2^i, r_3, r_4^i)^T,
\label{e:mev}
\ee
 ($m = f,s$), where
\bea
  r_1  & = &  \rho W^3 \Big(\rho h a (G + a^2 ) - G \mathcal{B}^2 / a \Big) ,
\nonumber \\
  r_2^i & = & W \Big( G \mathcal{B} B^i + \rho h W a^2 (\lambda_{m_\pm} v^i -
\zeta^i) \Big), \nonumber \\
  r_3 & = & r_1 p / \rho^2, \nonumber \\
  r_4^i & = & \rho h W^3 a \Big((\lambda_{m_\pm} \, v^i - \zeta^i)
\zeta_k B^k - B^i (\lambda_{m_\pm} a W^{-1} - G)\Big).
\eea
The scalar products are
\bea
{\cal P}_{m_{\pm}}^* & = & \Big(\frac{\partial \lambda_{m_\pm}}{\partial
\rho}\Big)\, r_1 + \Big(\frac{\partial \lambda_{m_\pm}}{\partial v^i}\Big)\, r_2^i
+ \Big(\frac{\partial \lambda_{m_\pm}}{\partial \epsilon}\Big)\, r_3 +
\Big(\frac{\partial \lambda_{m_\pm}}{\partial B^i}\Big)\, r_4^i \nonumber \\
 & = & \frac{W^3 a^4 G^2}{2 a_s^2 d} {\cal P}^*_1 \,{\cal P}^*_2,
\label{e:pm}
\eea
where $d$, the derivative of $\mathcal{N}_4$ with respect to $\lambda$
at $\lambda = \lambda_{m, \pm}$ ($m = f, s$), $\mathcal{N}_4'(\lambda_{m, \pm}) $, is

\be
d = a_s^2 G^2 \mathcal{B} (\zeta_k B^k)
- (G-\lambda_{m_\pm} a W^{-1})  \rho h W W_s^{-2} a^4,
\label{e:d}
\ee

and

\be
  {\cal P}^*_1 = b^2 G - \rho h a^2,
\label{e:p1}
\ee
\bea
\hspace{-2.5cm} {\cal P}^*_2 = \left(\rho \left. \frac{\partial
a_s^2}{\partial \rho}\right|_\epsilon + \frac{p}{\rho} \left. \frac{\partial
a_s^2}{\partial \epsilon}\right|_\rho\right)
W_s^2 \left(\frac{\mathcal{B}^2}{ a^2} - {\cal E}\right) - b^2 (3-a_s^2)
- 2 \rho h a_s^2 + \frac{a_s^2 (5 - 3a_s^2) \mathcal{B}^2}{a^2}.  & & \nonumber \\
 & &
\label{e:p2}
\eea

  It is interesting to note that $d$ can only be zero in degenerate
states, since it is only in these states where both 
$\mathcal{N}_4 (\lambda) = 0$ and $\mathcal{N}_4' (\lambda) = 0$ 
are satisfied
simultaneously. Let us now discuss the conditions under which the
remaining factors in Eq.~(\ref{e:pm}) can become zero. Quantity $a$ is
non-zero as far as $\zeta_k B^k \ne 0$. On the other hand, it can be
proven by simple algebraic manipulation of Eqs. $\mathcal{A}(\lambda)
= 0$ and $\mathcal{N}_4(\lambda) = 0$ that ${\cal P}^*_1 = 0$ if and only if the
corresponding magnetosonic eigenvalue is also an Alfv\'en eigenvalue
(i.e., Type II degeneracy). Since we are avoiding degenerate states,
and $G$ is always non-zero, we shall concentrate on the changes of sign
of ${\cal P}^*_2$, in order to analyze the possible loss of convexity
associated with the magnetosonic waves. 

  Since in the case of zero magnetic field, the purely relativistic
result has to be recovered, we shall try now to rewrite
expression (\ref{e:p2}) in terms of the relativistic fundamental
derivative
\be
\tilde{\cal G} =  1+ \frac{\rho}{2 a_s^2} \left. \frac{\partial
    a_s^2}{\partial \rho} \right|_s - a_s^2
\label{Gtilde}
\ee
derived in ref.~\cite{Ibanez13}. The sought expression is
\be
{\cal P}^*_2 =   - 2 a_s^2 W_s^2 {\cal E} (1 - R) \,
{\tilde{\cal G}_{\rm M}},
\label{e:p2new}
\ee
with ${\tilde{\cal G}_{\rm M}}$, the fundamental derivative for
relativistic, magnetized fluids, being
\be
{\tilde{\cal G}_{\rm M}} := { \tilde{\cal G}} +  F,
\label{C-RMHD-4}
\ee
where
\be
F:= \displaystyle{\frac{3}{2} W_s^{-4} 
\left(\frac{\displaystyle{c_a^2/a_s^2 - R}}{1 - R}\right)}.
\label{F-RMHD}
\ee
In the previous expressions, 
$R := \displaystyle{\frac{\mathcal{B}^2}{ {\cal E} a^2}}$, 
and $c_a^2 := \displaystyle{\frac{b^2}{{\cal E}}}$ 
stands for the square of the Alfv\'en velocity. Moreover,
in deriving expression (\ref{e:p2new}) from (\ref{e:p2}) we have used
the following relation among thermodynamical derivatives
$\displaystyle{\left. \frac{\partial \,\,\,}{\partial \rho}\right|_s =
  \left. \frac{\partial \,\,\,}{\partial \rho}\right|_\epsilon +
  \frac{p}{\rho^2} \left. \frac{\partial \,\,\,}{\partial
      \epsilon}\right|_\rho }$.

It is important to note that $R=1$ if and only if the eigenvalue 
corresponds to an Alfv\'en wavespeed (i.e., it satisfies
equation ~$\mathcal{A}(\lambda) = 0$). Since we are not considering 
degeneracies, we conclude that $R \ne 1$ for magnetosonic waves and, 
consequently, 1) the denominator in the second term of 
${\tilde{\cal G}_{\rm M}}$ is well defined, and 2) ${\cal P}^*_2 = 0$ 
if and only if ${\tilde{\cal G}_{\rm M}} = 0$. 

The price to pay for using primitive (or conserved) variables in 
our analysis of convexity is the loss of covariance and a dependence
of the fundamental derivative ${\tilde{\cal G}_{\rm M}}$ on kinematics
through quantity $R$. For fast and slow magnetosonic fields, let us
carry out the analysis of the magnetic correction to the purely 
hydrodynamic (relativistic) fundamental derivative (Eq. (\ref{F-RMHD})) in
the comoving frame (CF, $u^{\mu} = \delta_0^{\mu}$), which we will name 
$F_{{\rm CF},m}$ ($m= f, s$) henceforth. A simple algebraic calculation 
leads to 
\be
F_{{\rm CF}, m} = \frac{3}{2} W_\omega^{-2} \left(\frac{c_m^2 - a_s^2}{c_m^2 -
    c_a^2}\right),
\label{F1}
\ee
where $c_m^2$ are the solutions of the quadratic equation in $\lambda^2$,
$\mathcal{N}_{4, {\rm CF}} (\lambda) = 0$, namely
\be 
c_{m}^2 = \frac{1}{2} \left((\omega^2 + a_s^2 \, c_A^2) \pm 
\left((\omega^2 + a_s^2 \, c_A^2)^2 - 4 a_{s}^2 \, c_A^2 \right)^{1/2} \right), 
\label{N4LRF}
\ee
with $c_A^2 = \displaystyle{\frac{(\zeta_k B^k)^2}{\mathcal{E}}}$ and 
$W_\omega^{-2} := 1 - \omega^2$, $\omega^2 = a_s^2 + c_a^2 - a_s^2 c_a^2$. 

Taking into account that, for non-degenerate states, $a_s^2, c_a^2 \in
(c_s^2, c_f^2)$\footnote{In the CF it can be easily
  proven that $\mathcal{N}_{4, {\rm CF}} (a_s) < 0$ and
  $\mathcal{N}_{4, {\rm CF}} (c_a) < 0$, implying that both $a_s^2$ and
  $c_a^2$ are between the roots of $\mathcal{N}_{4, {\rm CF}} (\lambda)
  = 0$, namely $c_s^2$, $c_f^2$.}, we have that  $F_{{\rm
   CF}, m} > 0$ ($m = f, s$). Now, the transformation of $R$ as a
scalar ensures that $F_{m} > 0$ ($m = f,s$) in any
reference frame, with important consequences for the influence of the
magnetic field on the convexity of the system.

\end{itemize}

\section{Analysis of convexity in degenerate states} \label{s:cds}

\subsection{Type I degeneracy}

  This degeneracy appears in states in which $\zeta_k B^k = 0$. Now,
the roots of the characteristic equation~(\ref{caract}), the right
eigenvectors, and the corresponding scalar products have the
following properties:

\begin{itemize}

\item[i)] $\lambda = \lambda_{\rm null} := 0$. It is again the
  spurious eigenvalue analyzed in the previous Section associated with
  the null flux component. ${\mathcal P}_{\rm null}^*$ is trivially
  zero.

\item[ii)] The eigenvalue $\lambda = \lambda_0 := \zeta_k v^k$ has
  multiplicity 5. The corresponding eigenvectors are of the form ${\bf
r}^*_0 = (r_1, a_1 B^i + a_2 \zeta_\bot^i, r_3, a_3 B^i + a_4 \zeta_\bot^i)^T$,
where $\zeta_\bot^i$ is an arbitrary vector orthogonal to $\zeta^i$
and $B^i$, and $r_1$, $r_3$ and $a_p$, ($p=1,2,3,4$) are functions of
the primitive variables. Since only the derivative $\partial \lambda/ \partial
v^k$ ($= \zeta_k$) is different from zero, the scalar product is
\be
  {\cal P}_{0}^*  = \zeta_i (a_1 B^i + a_2 \zeta_\bot^i) = 0.
\ee
Hence, the characteristic fields defined by $\lambda_0$ are linearly
degenerate.

\item[iii)] $\lambda_{f,\pm}$ are the solutions of the quadratic equation
  in $\lambda$
\be
    \left(b^2 + \rho h a_s^2 - a_s^2 (v_k B^k)^2\right) G - W_s^{-2} \rho h a^2 = 0,
\label{e:char_deg_i}
\ee
and are associated with the fast magnetosonic wavespeeds. The explicit
expression of these eigenvalues when $\zeta_k = (1,0,0)$ can be found
in ref.~\cite{Leismann05}.

  The corresponding eigenvectors can be obtained from those of the
fast magnetosonic eigenvalues in the general case (see
Eq.~(\ref{e:mev})) making $\zeta_kB^k = 0$, i.e., $\mathcal{B} =  a (v_kB^k$).

  The scalar products are \footnote{As in the non-degenerate case, for the scalar products
  ${\cal P}_{f_{\pm}}^*$, the partial derivatives of the corresponding eigenvalues with
  respect to the primitive variables, ${\bf V}$, have been computed by
  implicit derivation of the characteristic equation~(\ref{e:char_deg_i}).}

\be
{\cal P}_{f_{\pm}}^* = \frac{W_s^2 G^2}{2 \rho h } {\cal P}^*_1
\,{\cal P}^*_2,
\label{e:pm-degi}
\ee
where
\be
{\cal P}_1^* = \displaystyle{\frac{{\cal E}  - (v_k B^k)^2 }{1 - \zeta_k v^k} }
\ee
\bea
{\cal P}_2^* & = & \left( \rho \left. \frac{\partial a_s^2}{\partial
\rho}\right|_\epsilon
+ \frac{p}{\rho} \left. \frac{\partial a_s^2}{\partial \epsilon}\right|_\rho \right)
 W_s^2 \left((v_k B^k)^2 - {\cal E}\right)  \nonumber \\
  & & - b^2 (3-a_s^2) - 2 \rho h a_s^2 + a_s^2 (5 -
  3a_s^2) (v_k B^k)^2.
\eea
From Eq.~(\ref{e:b2}), $b^2 - (v_kB^k)^2 \ge 0$ and then ${\cal P}_1^*$ is
always positive. Hence the possible changes of sign of ${\cal
  P}_{f_{\pm}}^*$ coincide with those of ${\cal P}_2^*$. Let us note
that the expression for ${\cal P}_2^*$ coincides with that of the
general case (Eq.~\ref{e:p2}) making $\mathcal{B} = a (v_kB^k)$. Then,
proceeding in exactly the same way as in the general case we conclude
that the fundamental derivative for relativistic, magnetized fluids
for Type I degenerate states is
\be
{\tilde{\cal G}_{\rm M, deg \, I}} = { \tilde{\cal G}} + 
\displaystyle{\frac{3}{2} W_s^{-4}
  \left(\frac{\displaystyle{c_a^2/a_s^2 - R_{\rm deg \, I}}}{1 - R_{\rm deg \, I}}\right)},
\label{C-RMHD-4_deg_I}
\ee
where now, 
$R_{\rm deg \, I} = \displaystyle{\frac{(v_k B^k)^2}{{\cal E}}}$. 
As discussed in the non-degenerate case, $R_{\rm deg \, I} \ne 1$, and the
corresponding factor is $F_{\rm deg \, I} > 0$.

  The special case when $v_k B^k = 0$ is obtained by making $R_{\rm deg \, I} = 0$ in
the previous expression. The same result for this case is obtained through a purely
hydrodynamical approach (see Appendix in ref.~\cite{Romero05}) by building up
a thermodynamically consistent EOS incorporating the
effects of the magnetic field.

\end{itemize}

\subsection{Type II degeneracy}

  Now, $\zeta_k B^k\ne 0$ and, at least, one eigenvalue associated
with an Alfv\'en wave and an eigenvalue associated with a
magnetosonic wave are degenerated. Three cases are distinguished. In
cases 1 ($c_a > a_s$) and 2 ($c_a < a_s$) one fast or slow magnetosonic
eigenvalue, respectively, and an Alfv\'en eigenvalue are
degenerated. In these cases, as discussed in the previous Section, the
quantity ${\cal P}^*_1$ defined in Eq.~(\ref{e:p1}) is zero for the
degenerate eigenvalues and, hence, the corresponding characteristic
fields are linearly degenerate. When $c_a = a_s$ (case 3), an Alfv\'en
eigenvalue is degenerated with a pair (slow and fast) of magnetosonic
eigenvalues. Now, quantity $d$ defined in Eq.~(\ref{e:d}) is
also 0, and we have an indetermination in ${\cal P}_{m_{\pm}}^*$
(Eq.~\ref{e:pm}). In this case, we have checked that the dot product
of the magnetosonic eigenvectors associated with the degenerated
fields and the gradient of the Alfv\'en eigenvalue is zero, which
means that the degenerate characteristic field is again linearly
degenerate. 

\section{Purely hydrodynamical and classical limits} \label{s:limits}

  The purely (relativistic) hydrodynamical limit can be obtained as a
particular case of the Type I degeneracy, in which besides having
$\zeta_k B^k = 0$ and $v_k B^k = 0$, we make $b^2 = 0$. Hence, from
Eq.~(\ref{C-RMHD-4}), and making $R = 0$ and $c_a = 0$, we have
${\tilde{\cal G}_{{\rm M}, b^2=0}} = { \tilde{\cal G}}$. 

  We now discuss the classical (magnetized) limits for both degenerate
and non-degenerate states. These limits are obtained by expanding all
the quantities in the definition of the fundamental derivative in
powers of $1/c^2$ ($c$ is the speed of light) and keeping the
leading term. On one hand, the relativistic (non-magnetized)
fundamental derivative is ${\tilde{\cal G}} = {\cal G} +
\mathcal{O}(1/c^2)$, where ${\cal G}$ is the classical
(non-magnetized) counterpart~\cite{Ibanez13}. On the other hand, 
$R = (\zeta_k B^k)^2/(\rho c_{m, {\rm cl}}^2) + \mathcal{O}(1/c^2)$,
where $c_{m, {\rm cl}}$ ($m=f,s$) is $c_{m, {\rm cl}} =
\displaystyle{\frac{1}{\sqrt{2}}} \left(a_{s, {\rm cl}}^2 + 
  B^2/\rho \pm \sqrt{(a_{s, {\rm cl}}^2 + B^2/\rho)^2 - 4 a_{s, {\rm
        cl}}^2 (\zeta_k B^k)^2/\rho} \right)^{1/2}$, and $a_{s,
  {\rm cl}}$ stands for the classical definition of the sound
speed. Hence, we get from Eq.~(\ref{C-RMHD-4})
\be
{\tilde{\cal G}_{\rm M, cl}} := {\cal G} +
\displaystyle{\frac{3}{2} 
\left(\frac{\displaystyle{c_{a, {\rm cl}}^2/a_{s, {\rm cl}}^2 -
    (\zeta_k B^k)^2/(\rho c_m^2)}}{1 - (\zeta_k B^k)^2/(\rho
    c_m^2)}\right)}. 
\label{C-RMHD-4_cl}
\ee
In the previous expression, $c_{a, {\rm cl}}$ stand for the classical
definition of the Alfv\'en speed, $\sqrt{B^2/\rho}$.

  It can be shown that, taking $\zeta_k = (1,0,0)$, the resulting
expression of ${\tilde{\cal G}_{\rm M, cl}}$ is proportional to the
non-linearity factor for the non-linear fields of the (classical) MHD
system obtained in ref.~\cite{Serna14} (see their equation~(17)).

  For Type I degenerate states, since $R = \mathcal{O}(1/c^2)$,
\be
{\tilde{\cal G}_{\rm M, deg \, I, cl}} = {\cal G} + 
\displaystyle{\frac{3}{2} \left(\frac{c_{a, {\rm cl}}^2}{a_{s, {\rm cl}}^2}\right)},
\label{C-RMHD-4_deg_Ib}
\ee
proportional to the corresponding result obtained in
ref.~\cite{Serna14} (see their table~I).

  Finally for Type II degenerate states, the eigenvalues that are
degenerated lead to characteristic fields which are linearly
degenerate, whereas the (hypothetical) non-degenerate magnetosonic
field (subcases 1 and 2) is genuinely non-linear and its properties in
relation with convexity are governed by the fundamental derivative in
Eq.~(\ref{C-RMHD-4_cl}), with $c_{m, {\rm cl}} = c_{s, {\rm cl}}$
(subcase 1), and $c_{m, {\rm cl}} = c_{f, {\rm cl}}$ (subcase 2).

\section{Summary and conclusions} \label{s:concl}

  In this paper we have analyzed the influence of the magnetic field
in the convexity properties of the RMHD equations. To this purpose we 
have used the approach of Lax, based on the analysis of the linearly
degenerate/genuinely non-linear nature of the characteristic
fields. Degenerate and non-degenerate states have been discussed
separately and the non-relativistic, unmagnetized limits are properly
recovered. The characteristic fields corresponding to the
material and Alfv\'en waves are linearly degenerate and, then, not
affected by the convexity issue. The analysis of the characteristic
fields associated with the magnetosonic waves reveals, however, a
dependence of the convexity condition on the magnetic field. 

  The result is expressed in the form of a generalized fundamental
derivative, Eq.~(\ref{C-RMHD-4}), written as the sum of two terms. The
first one is the generalized fundamental derivative in the case of
purely hydrodynamical (relativistic) flow already obtained in
ref.~\cite{Ibanez13}. The second one contains the effects of the
magnetic field. The analysis of this term in the comoving frame
(extendable to any other reference system given the scalar nature of
the term) shows that it is always positive leading to the remarkable
result that the presence of a magnetic field in the fluid reduces the
domain of thermodynamical states for which the EOS is non-convex, 
as it happens in the non-relativistic MHD limit~\cite{Serna14}.

We speculate with the possibility that our findings can be
relevant in the context of massive stellar core collapse. Depending
mostly on the pre-collapse stellar magnetic field and on the
gradient of the rotational velocity, dynamically relevant magnetic
fields may develop after the core bounce (see, e.g.,
\cite{Akiyama03,Obergaulinger06,Sawai13}). Should these magnetic
fields become as large as the existing numerical models point out,
then our results indicate that the loss of convexity would be rather
limited, if existing at all. However, it is still a matter of debate
what is the actual level of magnetic field saturation due to the
action of the Magneto Rotational Instability (MRI; see, e.g.,
\cite{Obergaulinger09,Pessah10}), and hence, whether or not the
MRI-amplified magnetic field may have the sufficient strength as to
impede the development of non-convex regions in the collapsed
core. It is very likely that under the most common conditions
(namely, non-rotating or slowly rotating cores), the magnetic field
will not play central dynamical role in the post-collapse evolution,
though it may set the time scale for supernova explosions (e.g.,
\cite{Obergaulinger14}). In such cases, we foresee that there might 
exist a range of physical conditions in which a non-convex EOS may 
render a convexity loss in the post-collapse core that cannot be 
compensated by the growth of pre-collapse magnetic fields, e.g., in 
slowly rotating (including non-rotating) massive stellar cores. 
Addressing this issue by means of numerical simulations is beyond the 
scope of the present work, and will be considered elsewhere.

\ack 
Authors acknowledge financial support from the Spanish Government
(grants AYA2013-40979-P and AYA2013-42184-P) and from the local 
Autonomous Government (Generalitat Valenciana, grant Prometeo-II/2014/069).
I. C.-C.  acknowledges support from the SN2NS project ANR-10-BLAN-0503, 
the ARC convention No. 11/15-040, and the Fonds de la Recherche Scientifique 
(FNRS) under grant 4.4501.05. M.A.A. and I.C.-C. acknowledge support from 
the European Research Councill (ERC) through the Starting Independent 
Researcher Grant CAMAP-259276.

\newpage

\noindent
{\bf Appendix. Jacobian matrices of the RMHD system in
  quasi-linear form}

\noindent
Matrices ${\bf {\cal A}}^0$ and $\zeta_k {\bf {\cal A}}^k$ associated with the system~(\ref{e:system}) in quasilinear form are:

\bea
{\bf {\cal A}}^0 = 
\left( \begin{tabular}{cccc}
 $W$  & $\rho W^3 v_j$ & $0$ & $0_j$ \\
 $({\bf {\cal A}}^0)^{S^i}_\rho$ &  $({\bf {\cal A}}^0)^{S^i}_{v^j}$ & 
$({\bf {\cal A}}^0)^{S^i}_\epsilon$ & $({\bf {\cal A}}^0)^{S^i}_{B^j}$ \\
 $({\bf {\cal A}}^0)^\tau_\rho$ &  $({\bf {\cal A}}^0)^\tau_{v^j}$ &
$({\bf {\cal A}}^0)^\tau_\epsilon$ & $({\bf {\cal A}}^0)^\tau_{B^j}$ \\
$0^i$ & $0^i_j$ & $0^i$ & $\delta^i_j$
\end{tabular} \right), \nonumber
\eea
where
\bea
	({\bf {\cal A}}^0)^{S^i}_\rho &=& (1+\epsilon+\chi) W^2 v^i, \nonumber \\
	({\bf {\cal A}}^0)^{S^i}_{v^j} &=& B^i B_j + B^2 \delta^i_j + h W^2 (\delta^i_j + 2 W^2 v^i v_j), \nonumber \\
	({\bf {\cal A}}^0)^{S^i}_\epsilon &=& (\rho+\kappa) W^2 v^i, \nonumber \\
	({\bf {\cal A}}^0)^{S^i}_{B^j} &=& -\delta^i_j v_k B^k - B^i v_j + 2 v^i B_j, \nonumber \\
	({\bf {\cal A}}^0)^\tau_\rho &=& (1+\epsilon) W^2 - W + \chi (W^2-1), \nonumber \\
	({\bf {\cal A}}^0)^\tau_{v^j} &=& -B_j v_k B^k + v_j [B^2 + \rho W^3 (2 h W-1)], \nonumber \\
	({\bf {\cal A}}^0)^\tau_\epsilon &=& \rho W^2 + \kappa (W^2-1), \nonumber \\
	({\bf {\cal A}}^0)^\tau_{B^j} &=& -v_j v_k B^k + B_j (2-1/W^2). \nonumber
\eea

\bea
\zeta_k {\bf{\cal A}}^k = 
\left( \begin{tabular}{cccc}
$W \zeta_k v^k$	& $(\zeta_k {\bf{\cal A}}^k)^D_{v^j}$	& $0$ & $0_j$ \\ 
$(\zeta_k {\bf{\cal A}}^k)^{S^i}_\rho$ & $(\zeta_k {\bf{\cal A}}^k)^{S^i}_{v^j}$
& $(\zeta_k {\bf{\cal A}}^k)^{S^i}_\epsilon$ & $(\zeta_k {\bf{\cal A}}^k)^{S^i}_{B^j}$ \\
$(\zeta_k {\bf{\cal A}}^k)^\tau_\rho$ & $(\zeta_k {\bf{\cal A}}^k)^\tau_{v^j}$
& $(\zeta_k {\bf{\cal A}}^k)^\tau_\epsilon$ & $(\zeta_k {\bf{\cal A}}^k)^\tau_{B^j}$ \\
$0^i$ & $B^i \zeta_j - \delta^i_j \zeta_k B^k$ & $0^i$ & $\delta^i_j \zeta_k v^k - v^i \zeta_j$\end{tabular} \right), \nonumber
\eea
where
\bea
	(\zeta_k {\bf{\cal A}}^k)^D_{v^j} &=& \rho W (W^2 v_j \zeta_k v^k + \zeta_j), \nonumber \\
	(\zeta_k {\bf{\cal A}}^k)^{S^i}_\rho &=& (1+\epsilon+\chi) W^2 v^i \zeta_k v^k + \chi \zeta^i, \nonumber \\
	(\zeta_k {\bf{\cal A}}^k)^{S^i}_{v^j} &=& (\zeta_i B_j - \delta^i_j \zeta_l B^l) v_k B^k + B^2 (\delta^i_j \zeta_k v^k - \zeta^i v_j + v^i \zeta_j) \nonumber \\
	&&- B^i (\zeta_j v_k B^k - 2 v_j \zeta_k B^k + B_j \zeta_k v^k) - v^i B_j \zeta_k B^k \nonumber \\
	&&+ \rho h W^2 (\delta^i_j \zeta_k v^k + v^i \zeta_j + 2 W^2 v^i v_j \zeta_k v^k), \nonumber \\
	(\zeta_k {\bf{\cal A}}^k)^{S^i}_\epsilon &=& v^i (\rho+\kappa) W^2 \zeta_k v^k + \zeta^i \kappa, \nonumber \\
	(\zeta_k {\bf{\cal A}}^k)^{S^i}_{B^j} &=& \zeta^i v_j v_k B^k - \delta^i_j v_k B^k \zeta_l v^l - B^i v_j \zeta_k v^k \nonumber \\
	&&- v^i (\zeta_j v_k B^k + v_j \zeta_k B^k - 2 B_j \zeta_k v^k) - W^{-2}(B^i \zeta_j - \zeta^i B_j + \delta^i_j \zeta_k B^k), \nonumber \\
	(\zeta_k {\bf{\cal A}}^k)^\tau_\rho &=& (1+\epsilon+\chi) W^2 \zeta_k v^k - W \zeta_k v^k, \nonumber \\
	(\zeta_k {\bf{\cal A}}^k)^\tau_{v^j} &=& -B_j \zeta_k B^k + B^2 \zeta_j + \rho W [\zeta_j (h W-1) + v_j \zeta_k v^k W^2 (2 h W-1)], \nonumber \\
	(\zeta_k {\bf{\cal A}}^k)^\tau_\epsilon &=& (\rho+\kappa) W^2 \zeta_k v^k, \nonumber \\
	(\zeta_k {\bf{\cal A}}^k)^\tau_{B^j} &=& 2 B_j \zeta_k v^k - v_j \zeta_k B^k - \zeta_j v_k B^k. \nonumber
\eea

  All the quantities appearing in the definition of the matrices are
defined in the body of the paper and $0^i = (0,0,0)^T$, $0_j =
(0,0,0)$ and $0^i_j$ is the null $3 \times 3$ matrix.
\newpage

\section*{References}

\end{document}